# Using administrative data to improve the estimation of immigration to local areas in England

Peter Boden
*University of Leeds, UK*
Phil Rees
*University of Leeds, UK*

**Summary.** International migration is now a significant driver of population change across Europe but the methods available to estimate its true impact upon sub-national areas remain inconsistent, constrained by inadequate systems of measurement and data capture. In the absence of a population register for England, official statistics on immigration and emigration are derived from a combination of survey and census sources. This paper demonstrates how administrative data systems such as those which capture registrations of recent migrants with a local doctor, National Insurance Number registrations by workers from abroad and the registration of foreign students for higher education, can provide data to better understand patterns and trends in international migration. The paper proposes a model for the estimation of immigration at a local level, integrating existing national estimates from the Office for National Statistics with data from these administrative sources. The model attempts to circumvent conceptual differences between datasets through the use of proportional distributions rather than absolute migrant counts in the estimation process. The model methodology and the results it produces provide alternative estimates of immigration for consideration by the Office for National Statistics as it develops its own programme of improvement to sub-national migration statistics.

*Address for correspondence:* Peter Boden, School of Geography, University of Leeds, Leeds, LS2 9JT, UK. E-mail: P.Boden@leeds.ac.uk

# Using administrative data to improve the estimation of immigration to local areas in England

## 1. Introduction

In 2004, the European Union (EU) was subject to significant expansion, with the accession of eight new member states from Central and Eastern Europe plus Malta and Cyprus. In 2006, further growth resulted from the integration of Bulgaria and Romania. The unprecedented inflow of economic migrants that resulted has left few parts of the UK and its economy unaffected by the impact of international migration, supplementing the historical flow of migrants from the New and Old Commonwealth, existing EU member states and other foreign countries (Bauere *et al.*, 2007).

A net inflow of new migrants, both from within and outside the EU has become a feature of demographic change across the UK and a key driver of population growth. National projections for the UK for 2006-11 estimate an annual population increase of 435,000, with approximately 50% due to natural increase and 50% due to net immigration (ONS, 2008a).

As the potential impact of this sustained net inflow of international migrants upon the UK's population has become apparent there has been increasing political pressure to regulate immigration. In early 2008 the Home Office introduced the first phase of the UK's new immigration system designed to simplify the process by which migrants from outside the European Economic Area (EEA) come to the UK (Home Office, 2006). The new points-based system (PBS) consists of five separate tiers, depending upon the level of skills offered by an individual, with each tier subject to different conditions, entitlements and entry-clearance checks. Points are awarded to migrants reflecting their skills, experience and age and the demand for these skills in the UK economy.

Yet despite the increasing importance of international migration to the UK economy, its impact upon local communities and the implementation of the new PBS, there remains no single source of statistics that provides a comprehensive measure of the new migrant population and its many dimensions. Since 2004 there has been much debate on the need for improved intelligence on the volume, profile and geographical distribution of migrant communities (Statistics Commission, 2007; House of Lords Select Committee on Economic Affairs, 2008a&b; House of Commons Treasury Committee, 2008; UKSA, 2009). In addition, in the absence of definitive statistics, regional and local bodies have been encouraged to create their own sources of intelligence on migrant populations, using local surveys, information from employers and recruitment agencies and through the sharing of data with service providers such as the police and health service (Audit Commission, 2007)

The necessity for improvements in migration statistics is not an issue that is confined to the UK, with an increased international policy focus on the socio-economic impacts of demographic change. As a dominant driver of population growth, robust estimation of immigration and emigration is key to the production of consistent national and regional population projections for EU countries (Lanzieri, 2007) and the implementation of EU Regulation 862/2007 has provided a statutory basis for greater harmonisation of international migration statistics in Europe. However, although the Nordic countries with their population registers provide the most accurate statistics, harmonisation of data collection remains a long way from being realised (Poulain et al., 2006) with significant issues of coverage and inconsistency between reporting countries (Kupiszewska and Nowok, 2005). Eurostat research, that has attempted to build Europe-wide international migration flow matrices for harmonised reporting and projections, has further revealed the differences in data quality between reporting countries and has shown the use of UK data, even at a national level, to be particularly problematic due to its small sample survey base (Raymer, 2008; Raymer & Abel, 2008).

Creative use of administrative data systems can provide a solution to the problem of migration estimation. In Canada, for example migration statistics are derived from address information recorded by the Canada Revenue Agency's database of annual tax returns (Statistics Canada, 1998), whereas in Australia net overseas migration is derived from an analysis of data captured directly from incoming and outgoing passenger cards (Australian Bureau of Statistics, 2006). In the absence of a population register in the UK and with census data ageing rapidly between successive enumerations, official estimates of immigration and emigration relies heavily on data collected through the International Passenger Survey (IPS). Administrative sources such as those operated by the Department of Works and Pensions (DWP) to allocate National Insurance Numbers (NINo) or by the Department of Health (DoH) to record registration with General Practitioners (GPs) are now being given serious consideration as alternative inputs to the estimation process (ONS, 2009).

The aims of this paper are: (1) to review the alternative sources of data available to measure international migration in the UK, (2) to summarise how some of these data are used by ONS to produce its sub-national immigration and emigration estimates, (3) to illustrate how a comparison of the time series of flows drawn from different sources helps our understanding of immigration at local scale and (4) to explore how information from the different data sources can be combined to produce a better, synthetic estimate of local immigration.

## 2. Alternative data sources

The pattern of passenger journeys is complex, with visitors and migrants coming into and out of the UK for a variety of reasons and for a variety of lengths of stay. Most visitors and migrants will enter the UK legally, though some will arrive as illegal migrants. Some come to find work, some to study, others to join existing family members and some to seek protection from abuse or persecution in their home country (asylum seekers). Some migrants will come with dependents. Some will come to the UK as visitors and then decide to stay for a longer period, sometimes for more than 12 months (visitor switchers). Others will come as migrants with the intention of staying for a long period but then change their mind and return within 12 months (migrant switchers).

Some migrants will be highly skilled, others less so, seeking manual and semi-skilled employment. Some will emigrate permanently from the UK; others will leave for a short or extended length of time but then return. Some migrants will come to the UK, stay in one place for only a short period but then move on to a more permanent residence.

The length of time a migrant stays in the UK is a particular issue when interpreting migration statistics. Those staying for less than three months, for example, are generally classed as visitors. A 'short-term migrant' is defined as, 'a person who moves to a country other than that of his or her usual residence for a period of at least 3 months but less than a year, except in cases where the movement to that country is for purposes of recreation, holiday, visits to friends and relatives, business, medical treatment or religious pilgrimage' (United Nations Statistics Division, 2006). At present, data on short-term migrants are not included in published statistics, although experimental data has been produced using additional intelligence from the IPS (ONS, 2008b) and ONS plans to release its first set of short-term migration statistics at local authority level during 2009.

A long-term migrant is a 'person who moves to a country other than that of his or her usual residence for a period of at least a year (12 months), so that the country of destination effectively becomes his or her new country of usual residence' (United Nations Statistics

Division, 2006). It is this migrant definition that is the basis for the UK's National Statistics on total annual immigration and emigration via a question on intended duration of stay in the IPS.

The distinction between stocks and flows is also an important one when using migrant statistics. Stocks provide a count of the total number of resident migrants. Resident migrants can be counted in a number of different ways: as persons with a different country of birth or as persons who have immigrated to the UK within a specified time period. Flows provide a count of the number of new migrants that come to or leave the UK in a specified period of time, usually a single year. Migrant flows will increase or decrease the size of the resident stock of migrants, depending upon the balance of emigration to immigration.

Migrant flows can be measured in one of several ways. The number of border crossings or migration events can be counted, as is intended to be the case when electronic passports are in full use. Within a time interval a migrant can make several migrations, some or all of which cancel out in net terms. The main alternative method of measuring migration is to compare a person's location at two successive points in time, 12 months apart in the case of long-term migrants. A comparison of locations determines whether a migration took place. In the IPS the comparison is between the time of entry and 12 months in the future, thus measuring intentions. The census, by contrast, asks respondents about their locations 12 months ago and so measures what has happened, at least to survivors. Both measures miss migration events and are affected by deaths to migrants.

There is no single data collection instrument for the measurement of international migration. There are a number of alternative sources which provide specific intelligence about the movement of population into and out of the UK. These sources may be generally classified as either census, survey or administrative datasets. Each has its own limitations depending upon the question asked, the purpose of the data collection and the population covered (Table 1).

[Table 1 about here]

Table 1 is structured as follows. The sources occupy different rows of the table, grouped into three categories – censuses, surveys and administrative datasets. The columns provide information on the attributes of the sources organised into three groups – source features, coverage attributes and definitional characteristics.

The decennial census is the most comprehensive source of data on the UK population but its data ages rapidly, particularly at a time of such significant demographic change. Surveys are rich sources of data but are typically not statistically robust for local-area analysis and do not adequately capture all migrant populations. Administrative sources can provide excellent geographical detail but typically do not have the data richness that a survey provides. Few sources provide data on emigration from the UK with administrative systems typically only providing data on new or resident migrants.

A more detailed description of the content and quality of alternative sources of data can be found elsewhere (Rees and Boden, 2006, Green *et al*, 2008). A short summary of each is provided here using the three-fold classification of censuses, surveys and administrative datasets.

**2.1 Censuses**

The **Census** can provide both a view of migrant flows for the year prior to enumeration and a measure of migrant stocks present in the usually resident population. Migration 'flow' data are derived from a question, which asks for an individual's address twelve months prior to enumeration day. Only in-migration is measured as there is no attempt to capture information on individuals who have emigrated during the Census year. Some censuses do ask households to report members who have recently emigrated, but the emigrations of wholly moving households are missed. The stock picture is derived from detailed country of birth statistics, although in the absence of a question on year of entry to the UK, it is not possible to measure the length of time a migrant has been resident in the UK. In 2011 the Census will ask the question on *usual residence one year prior to census date* (asked since 1961) and the question on *country of birth* (asked since 1851) (ONS 2008d). New questions will be asked on *month and year of entry to the UK* and *intended length of stay in the UK* for all those born outside the UK and on *citizenship*. Census migration statistics are available for download from the NOMIS website (www.nomisweb.co.uk/Default.asp).

The **Pupil Census** or National Pupil Dataset (NPD) contains individual pupil records for all children in grant maintained schools in the UK. The dataset is managed by the Department for Children, Schools and Families (DCSF) and collected on a twice-yearly basis from individual schools within each Local Education Authority (LEA). A total of 8 million pupils are included in the dataset each year. The dataset does not provide an obvious source of statistics on migrant flows but it does have the potential to provide an informed picture of the composition of local areas based on the changing profile of pupil numbers using information captured on ethnicity and first language. The complexities of data processing have so far precluded more significant use of the Pupil Census as a migration data source. It does not feature in the analysis presented here but more information and a number of data downloads may be obtained from the University of Bristol's User Group website (www.bris.ac.uk/Depts/CMPO/PLUG/whatisplug.htm).

**2.2 Surveys**

The **International Passenger Survey** (IPS) is the only instrument for measuring UK immigration and emigration, for nearly all types of migrant. It is a multi-dimensional survey, of which the migrant questions are just one part. It surveys approximately 250 thousand passengers each year: the sample members constitute about 1 in 400 of the total number entering or leaving at the UK ports. Of this sample, about 1% are migrants whose stated intention is to stay or leave the UK for more than 12 months. This is equivalent to approximately 3,000 respondents, 70% of whom are immigrants and 30% are emigrants, which need to be re-weighted to provide representative statistics. From 2007, the number of interviews with departing migrants has been boosted to a comparable level to those on entry. IPS statistics on long-term migration (where duration of stay is more than twelve months) feed directly into National Statistics of Total International Migration (TIM) produced by ONS (see section 3). IPS statistics are available for download from the Office for National Statistics website (www.statistics.gov.uk/statbase/product.asp?vlnk=15240).

The **Labour Force Survey** (LFS) is a quarterly sample survey of households living at private addresses in the UK and provides the most detailed statistics on the UK labour market. The LFS captures a 60,000 sample of households in Great Britain and asks the question 'where were you living one year ago', so it can provide a count of the 'flow' of migrants coming to the UK within a single year. It also records information on year of entry to the UK, which provides a picture of the length of time migrants have been resident – thus producing the most reliable statistics on the 'stock' of migrant workers in the UK. As an accurate measure of international migration the LFS has a number of drawbacks. It excluded students in halls of residence who do not have a UK resident parent plus it excluded people in most types of communal establishments and those migrants who have been in the UK for less than six months. In addition, LFS totals are grossed to population estimates that only include long-term migrants. The LFS contains a sample of about 700 international migrants per year (i.e. persons who state they were resident overseas one year ago). This small sample size precludes more detailed analysis of migrant inflow by local geographical area.

The LFS has been combined with the Annual Population Survey (APS), the General Household Survey (GHS), the Expenditure and Food Survey (EFS), and, the National Statistics Omnibus Survey (NSOS) to create the **Integrated Household Survey (IHS)**. This single survey approach has created a much larger sample size, with migrant worker questions from the LFS being retained in a core module that is expected to cover 221,000 households (ONS, 2007c). In addition, the LFS data capture is now more likely to pick up short-term migrants, with respondents interviewed at their current address regardless of how long they have lived there. Statistics from the LFS are accessible from the UK Data Archive website (www.data-archive.ac.uk/findingData/lfstitles.asp).

**2.3 Administrative datasets**

The **Home Office** regularly publishes National Statistics on immigration and asylum. British Citizens, those Commonwealth Citizens who have freedom of entry to the UK and nationals from the European Economic Area (EEA) are not subject to immigration control and are not included in Home Office statistics. No information is recorded on people emigrating from the UK. National Statistics produced by the Home Office fall into three broad categories: asylum seekers, persons subject to immigration control and persons granted British Citizenship (Home Office 2008a, 2008b). Most statistics are only available at a national level, with no sub-national provision. **Work Permit** statistics for each local authority district and unitary authority have previously been made available but these data are no longer routinely produced by the Home Office, although **asylum statistics** are still available at a local authority level. Aggregate immigration statistics are routinely published by the Home Office (www.homeoffice.gov.uk/rds/immigration-asylum-stats.html). Work permit data must be accessed through a freedom of information request.

For a new migrant to the UK, acquiring a **National Insurance Number** (NINo) is a necessary first step for employment/self-employment purposes or to claim benefits or tax credits. NINo statistics, managed by the Department of Works and Pensions (DWP), record an individual's residence, 'country-of-origin', age and gender. The Information Directorate (IFD) within DWP is responsible for the publication of statistics from its National Insurance Recording System (NIRS) and a summary of NINo registrations by A8 migrants is published periodically as part of the more general release of migration statistics coordinated by ONS (Home Office, 2008c). NINo statistics exclude dependents of applicants, unless they claim benefits or work themselves. They will also exclude most students and those migrants who are not of working age and not claiming benefits. Although year of arrival in the UK is recorded, the statistics provide no indication of the length-of-stay of a migrant worker and there is no formal de-registration process. Migrants can actually leave the UK and return at a later date without the necessity to re-register for a new NINo. NINo statistics are available for download from the DWP's online tabulation service (http://83.244.183.180/mgw/live/tabtool.html).

Until April 2009, nationals from the Accession 8 countries of the Czech Republic, Estonia, Hungary, Latvia, Lithuania, Poland, Slovakia and Slovenia who come to work in the UK were required to register with the **Workers Registration Scheme** (WRS). A new registration is required when a person changes employment or an applicant is employed by more than one employer. Year of registration is recorded, as is nationality of the individual. Date of birth, gender and occupational status are also routinely captured. There is no method for tracking how long each applicant stays in the UK as, like the NINo system, there is no de-registration process necessary. A detailed statistical picture produced from the WRS is regularly published, illustrating the profile of applicants and of registered workers and detailing type of employment, hours of work, wages and a regional disaggregation (Home Office, 2008c).

The WRS provides richer data detail than NINo statistics but has a number of limitations. It records the location of the employer but not the residential location of the applicant. It only records information on A8 migrants and it also excludes those who are self-employed. In addition, the WRS will not record A8 migrants who come to the UK for reasons other than work, including students. The WRS is also only a temporary administrative system and terminated in April 2009. WRS data is available through a Freedom of Information request to the Home Office.

When new migrants first register with a **General Practitioner** (GP), they are explicitly identified as an individual whose previous address was outside the UK and who has spent more than three months abroad (their records are labelled as 'Flag 4' data). The Patient Registration Database System (PRDS) records the age and gender of new migrants but does not provide any more detailed information on nationality, country of origin or country of birth. No information is captured on patients who have emigrated from the UK. GP registrations capture all migrants, regardless of age and employment status, so in theory they provide the most comprehensive view of migration inflows. Migrants captured by the registration process will include short-term migrants, in addition to those who have been resident for at least twelve months. It is not possible to identify actual or intended length of stay from the data. For the majority of migrants, there will be a time-lag between entering the UK and registering with a GP and some migrants may never complete the registration process during their stay in the UK. Young men, in particular, will delay registration after migration more than older men or women. Also, a PRDS record loses its migrant status once a patient moves within the UK and registers with a new GP. GP registration statistics are not routinely published and a request for use in research and analysis should be addressed directly to the Office for National Statistics.

The **Higher Education Statistics** Agency (HESA) maintains a record of all students in the UK, including those whose country of usual residence is outside the UK. HESA administrative systems do not capture the residential address of international students, only the location of the institution of study. Efforts are being made by HESA with ONS and other agencies and universities to record students by their term-time usual residence. Students provide information on their expected length of stay and although nationality is requested, it is not a mandatory field and coverage is typically poor. Age and gender are recorded and ethnicity is only provided for students with a UK domicile. A 'flow' picture can be produced, recording all students who arrive and depart in a particular year. In addition, by looking at all students who are studying during a particular year, a 'stock' picture can be produced. The picture is dynamic because of the constant churn of students by institution. HESA statistics are only available under licence. More information on usage is available from the HESA website (www.hesa.ac.uk/index.php/content/view/35/134/).

## 3. Official estimates: TIM methodology

In England and Wales, Total International Migration (TIM) statistics provide the most accurate estimates of **long-term** immigration and emigration at a national level (ONS, 2008c) and the basis for estimating the immigration and emigration components used in the production of mid-year population estimates and sub-national population projections. The TIM estimates of immigration and emigration are primarily based on sample data derived from the question asked in the IPS on migrants 'intentions' to stay in or leave the UK for more than twelve months. These are supplemented with additional statistics from the Home Office on asylum seekers and their dependants and from the Irish Central Statistical Office (ICSO) on estimates of migration between the UK and the Irish Republic based on the Quarterly National Household Survey (QNHS) in Ireland. Since 2004, estimates of visitor switchers (those people whose original intention was to stay for less than twelve months but who subsequently stay for longer) and migrant switchers (those people who intended to stay for more than twelve months but decide to leave within a year) have been derived from questions in the IPS (ONS, 2008c).

[Figure 1 about here]

The estimation process works at four different levels: national, regional, intermediate geography and local authority area. The intermediate geography or 'New Migrant geography' (NMG) groups local authority areas outside London based upon their contiguity and economic association. Within London, boroughs are grouped based on similarities in historical migration profiles. There is a different set of these new geographies for both immigration (NMGi) and emigration (NMGo). Detail on their design and definition is available elsewhere (ONS 2007a; ONS 2007b).

For immigration estimation the LFS is used to allocate gross national IPS flows to the nine Government Office Regions (GOR). LFS statistics on 'long-term' migrants calibrate the proportional distribution of flows to each region. IPS data, smoothed over an extended time-series, is used to allocate immigration flows to the NMGi outside London, with the LFS sample size believed to be sufficiently robust to enable it to be used for estimation for the London NMGi areas. The final stage of immigration estimation involves the proportional allocation of flows to local authority areas using the migrant distributions evident from 2001 Census data..

[Figure 2 about here]

The emigration estimation process has a similar hierarchical structure but does not have the luxury of additional data from either the LFS or the Census. As an alternative, it incorporates a 'migration propensity' model to estimate the distribution of flows at a local authority level. A more complete description of the TIM estimation methodology is provided in ONS supplementary documentation (ONS 2008c).

The principal innovation of the new ONS methodology is to use the LFS to re-weight regional distributions. Since the LFS is small sample comparable to the IPS with only circa 700 international immigrants per year and with small sub-samples for home countries and GORs, it is likely that the confidence limits around any estimate are very wide. The LFS is a household survey which does not cover communal establishments. Since about 25% of immigrants, according to the IPS, come to study, this is likely to lead to distortion of the sub-national distributions. The method of calibration of the IPS to the LFS is probably responsible for the increased allocation of immigrants to Yorkshire and the Humber (ONS 2007a, Figure 2). At the lowest level, from Intermediate Geography to Local Authority, ONS uses the distribution found in the 2001 Census. There have been considerable changes in the volume and distribution of immigration since 2001, so that this final step is likely to be in more and more error as the decade proceeds. The distribution of A8 immigrants, for example, is very different from the 2001 Census all immigrant distribution (Bauere *et al.* 2007).

The TIM estimation methodology is applied to all local authority areas in England and Wales. The remainder of this paper focuses on the local authority areas of England, for which a complete set of immigration statistics from all sources was available. Importantly, Scotland and Northern Ireland employ alternative methods for the sub-national estimation of immigration flows. The General Register Office Scotland uses the Community Health Index (GP registrations) to allocate national IPS totals to council areas and the Northern Ireland Statistics and Research Agency uses the same data to directly estimate its international migration flows (GROS, 2007a; NSCD, 2006).

To improve the estimation of immigration and to combat some of the issues identified above, the latest methodological developments from ONS have investigated the integration of administrative data as the basis for allocating flows from intermediate geographical zones to local authority areas (ONS, 2009). The research presented here is designed to complement these developments and to anticipate the issues and impacts that arise from the use of new sources of data for the estimation of sub-national immigration flows.

## 4. A comparison of immigration statistics

To facilitate the analysis of the patterns and trends in sub-national immigration and emigration, the New Migrant Databank (NMD) has been created as a unique repository of statistics on international migration in the UK (Boden and Rees, 2009). The NMD has harmonized on data for local authority areas for the 2001-2008 time-period, integrating census, survey and administrative sources to enable presentation of the patterns and trends that are evident and the analysis of differences that exist between datasets, particularly between TIM statistics and administrative sources.

The trend in immigration flows to England since 2001 are illustrated in Figure 3, based on three data sources with the census serving as a benchmark. In the year prior to the 2001 Census long-term immigration to England was recorded at approximately 361,000 migrants. According to TIM estimates long-term immigration has increased from 450,000 in 2001/02 (mid-year to mid-year intervals) to 533,000 by 2006/07, an 18% rise over five years. The registration of new migrants with a GP has followed a similar trend to TIM since 2001, although these registrations will include some migrants whose duration of stay was less than 12 months (although greater than 3 months as registration can only be completed after this length of stay in the UK). In 2001 there were 389,000 new GP registrations to foreign nationals, rising to 581,000 in 2007, an increase of almost 50% during the six-year time-series.

[Figure 3 about here]

NINo registrations paint a more dramatic picture of change over a five-year time-series. NINo statistics record all registrations, whether for long-term or short-term employment in the UK. They will, however, exclude those individuals who do not work, such as non-working partners, dependants and students. The statistics are split into 'non-Accession' and 'total' numbers to illustrate the impact of Accession migrants upon immigration flows. In 2002/03, prior to expansion of the EU, the vast majority of registrations were to non-Accession migrants: 294,000 in 2002/03, increasing to 350,000 in 2005/07 and then declining to 341,000 in 2007/08. The curve for 'all' NINo registrations illustrates the impact of Accession-country migrants, reaching 276,000 in 2007/08 giving a total number of registrations for England of 617,000.

At a regional level there are some notable differences in the trends evident from the data sources (Figure 4). GP registrations show evidence of consistent increases across the regions since 2004 with a slightly flatter trend in London compared to other areas. The relationship between GP registrations and TIM estimates of immigration, although reasonably consistent at a national level, shows more variation regionally – particularly in the West Midlands, Yorkshire and the Humber, South West and London.

[Figure 4 about here]

The national trend in NINo registrations is generally reflected throughout the regions although in the last year of the time-series, whereas most regions have evidence of a reduction in both 'total' and 'non-Accession' migrants, London registrations have continued to rise. This is likely evidence that economic downturn, which began in late 2007, has affected regions outside London, whereas the capital has, for the time being, continued to attract Accession migrants, bolstered by the impetus of new migrant flows from Bulgaria and Romania.

The national profile of immigration presented in Figure 3 illustrated that although TIM statistics and GP registrations are conceptually different measures of migration, the general level and trend of each data series for England is quite consistent. This suggests that GP registration statistics could provide a useful comparative measure of the level and distribution of long-term immigration estimated by the TIM process. Figure 5 compares aggregate TIM statistics and GP registration data for each of the English regions for the three-year period 2005-2007. Given the similarity in the national picture, one might expect general consistency between the two datasets at a regional level but the graph illustrates that there are significant differences evident, particularly for Yorkshire and the Humber, the South West and the West Midlands. In the South West and Yorkshire and the Humber, TIM estimates of immigration were, in aggregate, over 16% higher than the total number of GP registrations in the corresponding period. In the West Midlands they were 34% lower. These differences are emphasised in the respective time-series plots presented in Figure 4.

[Figure 5 about here]

An alternative comparison of TIM estimates and GP registrations is presented in Table 2, with immigration flows for each region converted to 'rates' of immigration in 2006, based upon the resident population of the receiving region in mid-year 2006. TIM statistics suggest an immigration rate of 10.5 per thousand for England, 8% lower than the corresponding GP registration rate of 11.5 per thousand. Comparing the relative rates of immigration across the regions indicates significant differences between the two, evidence possibly of differential preferences for particular regions but more likely to indicate the potential inconsistency of the TIM estimates between regions. The West Midlands, for example, has a rate of immigration from GP registrations which is 30% higher than the corresponding TIM rate, whereas in the East of England it is 10% higher.

[Table 2 about here]

There are clearly some inconsistencies that exist at a sub-national level between TIM statistics and immigration totals recorded in administrative sources. The following section introduces a method to enhance TIM sub-national estimates of immigration through the direct use of data from three administrative sources: GP registrations, NINo registration and HESA international student statistics.

## 5. Improving immigration estimates using administrative sources

Administrative data sources typically provide excellent geographical detail and usually capture a very large population 'sample'. By contrast, survey sources typically provide much greater data detail but from a limited sample size, precluding use at smaller spatial scales. The TIM estimation process uses a combination of survey and census sources. Importantly, it produces a measure of immigration and emigration for all types on long-term migration. Adminsitrative sources, such as NINo and GP registrations, cannot simply replace these estimates as their respective data typically only include a sub-set of the population (workers in the case of NINo) and does not distinguish the length of stay of migrants (both NINo and GP registrations).

This section presents an alternative process of estimation for immigration flows which retains the TIM estimates at a national level but uses the proportional distribution of immigration flows evident from administrative sources to allocate this total to sub-national areas. This process accepts that TIM estimates are robust at a national level but proposes that a combination of administrative sources provides a more appropriate method of allocation to smaller geographical areas than the existing approach based upon survey and out-dated census sources. The alternative estimation models that have been tested are illustrated in Figure 6.

[Figure 6 about here]

Model A experimented with the direct allocation of national TIM estimates using GP registration distributions for each local authority area. Given the different types of immigration flow (worker, student, other) and the availability of administrative sources recording these activities, it was felt that a more sophisticated allocation process could be deployed. At a national level it is possible to disaggregate TIM flows based upon a 'reason profile' (Figure 7a). For example, 26% of all long-term immigration flows between 2004 and 2006 were associated with 'formal study' and 27% were linked to a 'definite job'. Model B uses a combination of three data sources to allocate immigration flows to sub-national areas:

1. The distribution of NINo registrations to all England migrant workers was used to allocate the 40% of flows that were associated with migrants whose stated reason for migration was either a 'definite job' or 'looking for work'.

2. HESA statistics for 2005-2007 were aggregated by local authority of institution to identify the distribution of international students by local authority. The resulting data was then used to allocate the 26% of migrants who stated reason for migration was 'formal study'

3. Finally, GP registration statistics were used to allocate the remaining 34% of migrants whose stated reason for migration was either 'accompany/join', 'other' or 'not stated'.

[Figure 7 about here]

The HESA data is captured by institution and was therefore not perceived to be robust for sub-national allocation at anything below regional level, particularly in London where colleges operate from split sites and where place of study can be very different from place of residence. For this reason GP registrations were used to achieve the most geographically disaggregate distribution.

The specification of the allocation process used in Model B is as follows:

$$M_j = \left[ \sum_k M \, r_k \, q_{Jk} \right] s_{(j|J)} \tag{1}$$

where
- $j$ = local authority district
- $J$ = Government Office Region (GOR)
- $k$ = reason for immigration (1 formal study, 2 definite job or looking for work, 3 other)
- $M$ = Total International Migration (TIM) immigration estimate for the UK

$M_j$ = Immigration estimate by local authority district j

$r_k$ = $(M_{jk}/M)$ = TIM immigration proportion by migrant type k

$q_{jk}$ = $\dfrac{H_{Jk}}{\sum_J H_{Jk}}$ = the proportion of the administrative dataset count, $H$, for GOR J and migrant type k of UK total of migrant type k

$S_{(j|J)}$ = $\dfrac{\sum H_{3j}}{H_{3j/j \in J}}$ = the proportion of the GP registration count for local authority district j in GOR J, where, $H_{3J}$ = count of migrants of type 3 for GOR J and $H_{3j}$ = count of migrants of type 3 for local authority j

The results of Model 3 and its comparison to TIM estimates for 2006 are presented here.

The use of the three administrative sources to distribute TIM national estimates produces some significant differences in the regional estimates of immigration (Figure 8). London achieves the most significant increase, with an additional 20,334 migrants compared to the TIM estimates. The West Midlands also increases its share of the immigration pot by over 11,000 with smaller increases in the North West and the North East. In percentage terms, the increase in the West Midlands is largest, a 33% increase on the TIM estimates.

As this is a redistribution of flows based on the proportions evident in administrative datasets, those regions that gain are balanced by those that see a reduction in their immigration estimates. The East of England experiences the largest reduction both in absolute (-13,584) and percentage (-23%) terms. The South West and Yorkshire and the Humber also experience significant reductions (19% and 21% respectively) with smaller reductions in the South East and the East Midlands. The administrative datasets clearly suggest a distribution of new migrants that is significantly different from that evident in the TIM estimates.

[Figure 8 about here]

The second stage of the estimation process associated with Model B allocates immigration flows to sub-regional areas (the NMGi 'intermediate' geography adopted by ONS in its estimation methodology and local authority areas) based on the distribution of GP registrations. The West Midlands and Yorkshire and the Humber are used as illustrative examples, with the former gaining new migrants as a result of the new estimation process and the latter losing (Figure 9).

[Figure 9 about here]

In Yorkshire and the Humber the reduction of 10,292 immigration flows to the region is distributed using the GP registrations for the three constituent NMGi areas with the Leeds/Bradford/Harrogate+ area seeing a reduction of 7,644 (-30%), Sheffield/Kirklees+ 2,339 (-15%) and Hull/York+ a smaller 309 (-4%) (Figure 9). In the West Midlands the Staffordshire/Shropshire/Wolverhampton NMGi experiences a substantial 89% increase in its immigration total, compared to a 6% loss in the more rural Herefordshire/Worcestershire/ Warwickshire NMGi. The Birmingham and Coventry NMGi both experience increases of over 30% in their immigration estimates.

Despite losses to each NMGi in Yorkshire and the Humber in the Model B estimates compared with the TIM estimates, the allocation process does produce both losses and gains to individual local authority areas within the region (Figure 10).

[Figure 10 about here]

North Lincolnshire, Wakefield and Selby experience the largest percentage gains. In South Yorkshire, Rotherham, Doncaster and Barnsley all have marginal gains, whereas Sheffield has a 29% reduction in its immigration flow total. The largest percentage reductions are associated with small absolute changes in the rural authorities of North Yorkshire. The largest overall reduction is in Leeds, the economic focus of the region, losing almost 5,000 from its TIM immigration estimate, a 36% fall.

In the West Midlands the most significant increases in the estimate of immigration flows are in Birmingham (2,931), Sandwell (1,679), Stoke-on-Trent (1,250), Coventry (1,226) and Wolverhampton (1,234) (Figure 11). In percentage terms, East Staffordshire increases its immigration total by 255%, Telford and Wrekin by 188%. In contrast to these gains are losses to a number of local authority areas. Warwick and North Warwickshire estimates are of particular interest as they will each be influenced by the presence of international students studying at Warwick University which is actually located in the Coventry local authority. This is an example of where further model refinement through local consultation would help to improve the estimation of immigration flows at a local level.

[Figure 11 about here]

A process of local consultation may be an appropriate strategy for a number of local authorities where the impact of the Model B allocation procedure is most significant. Figure 12 presents a list of the top 20 'gainers' and 'losers', reflecting the difference between the TIM estimate and the new Model B estimate of immigration.

[Figure 12 about here]

In addition to the changes evident in the West Midlands and Yorkshire and the Humber (Birmingham has the second largest increase in immigration flows; Leeds the third largest decrease) there is a significant redistribution of immigration flows in London, with major losses in Westminster, Kensington and Chelsea and Camden balanced by large gains in a number of boroughs, particularly, Newham, Wandsworth, Lambeth, Hounslow, Greenwich, Ealing and Waltham Forest. There are other local authorities where the scale of the change requires further investigation to identify any local factors which might be influencing the significant differences between TIM and the new immigration estimates. The impact upon Cambridge, for example, needs to be scrutinised to assess the effect of student distributions in adjacent local authorities. The large increase in the immigration estimate for Peterborough should be verified against local evidence and knowledge.

## 6. Future development

International migration is now a dominant driver of sub-national population change across Europe, yet there remain significant shortcomings in our ability to accurately measure and estimate the true impact of immigration and emigration flows. This paper has used English local authority areas to illustrate how existing estimation methods employed by the UK's Office for National Statistics could be enhanced through the integration of proxy measures of migration derived from administrative sources.

The New Migrant Databank is a unique repository of UK statistics on international migration, harmonized at the local authority level, for an extended time-period, 2001-2008. It has provided the empirical base for the analysis of international migration at a sub-national level, highlighting the issues and difficulties associated with the derivation of accurate estimates of immigration and emigration. The analysis presented here has shown that there are major dissimilarities between the pattern of immigration evident from ONS local area estimates derived from its TIM methodology, and that derived from administrative sources, specifically GP registration statistics.

The paper proposes an alternative estimation model which combines TIM estimates at a national level with data from three administrative datasets: HESA, NINo and GP registrations. The approach presented is not perfect but innovative in that it uses 'local' and 'current' datasets where available, as an alternative to the relatively small survey samples (IPS and LFS) and increasingly out-of-date census information. The model attempts to circumvent the definitional differences between datasets by using the proportional distribution of administrative counts rather than actual totals as the basis for local area estimation. In addition, it is likely that the direct use of administrative data will better capture 'first onward move' of recent migrants avoiding the 'temporary' or 'false' locations that may be recorded at initial IPS data capture.

Refinements to the model will continue to be made through scrutiny of local factors that may have a particular influence upon the distribution of the respective administrative statistics and thus on the immigration estimates produced. Further sources, such as the Pupil Census which has yet to be added to the New Migrant Databank, will provide additional local intelligence on new migrant activity. The compatability of the different sources remains an issue as each administrative system captures a different population sub-group. Record linkage between datasets using a unique person reference number (NINo or NHS number, for example) would enhance the combined value of the data but this would require major investment and would be subject to inevitable Data Protection constraints. The use of covariates to estimate immigration flows at a local level is an alternative approach that could remove potential biases evident in administrative sources and is a method now being investigated by ONS as part of its ongoing improvement programme for the UK's population statistics (ONS, 2009).

Of course the international migration picture has two key dimensions – immigration and emigration. The estimation of emigration flows by local authority area is even more problematic. Census and survey sources do not capture emigration and administrative sources do not regularly record 'de-registration' activities associated with migrants leaving the UK. For this reason, the IPS remains the only reliable source of data on emigration from the UK and ONS uses a 'propensity-to-migrate' model to estimate flows at a local level.

Furthermore, the impact of short-term migration has become increasingly evident in local areas for the UK since the expansion of the EU in 2004. Those migrants staying in the UK for less than twelve months have become an important component of local populations. Population estimates ignore this sub-population and there is a dearth of reliable statistics on the average length of stay of migrants, either from survey or administrative sources. However, experimental statistics have been published by ONS using data from the IPS on migrant intentions and more comprehensive estimates by local authority area are due for publication during Autumn 2009.

There remains a substantial research agenda in the field of international migration measurement and, in the absence of a population register, it is essential that all data sources, survey, census and administrative, are used to their full potential to ensure the most robust evidence is available to support international, national, regional and local policy development.

**Acknowledgements**

The authors gratefully acknowledge the support of ESRC through award RES-163-25-0032.

Project title: 'What happens when international migrants settle? Ethnic group projections for UK local areas.' The project is part of ESRC's *Understanding Population Trends and Processes* (UPTAP) programme, directed by John Stillwell.

The authors also wish to acknowledge those organisations which provided data to the research project and that have been used in the analyses presented here: the Department of Works and Pensions, the Office for National Statistics and the Higher Education Statistical Agency. These organisations do not accept responsibility for any inferences or conclusions derived from the data analysis presented in this paper..

## Glossary

| | |
|---|---|
| A8 | Accession 8 countries: Czech Republic, Estonia, Hungary, Latvia, Lithuania, Poland, Slovakia and Slovenia |
| Accession | A8 plus Bulgaria and Romania |
| APS | Annual Population Survey |
| DCSF | Department for Children, Schools and Family |
| DoH | Department of Health |
| DWP | Department of Works and Pensions |
| EEA | European Economic Area |
| EFS | Expenditure and Food Survey |
| ESRC | Economic and Social Research Council |
| EU | European Union |
| GHS | General Household Survey |
| GLA | Greater London Authority |
| GOR | Government Office Region |
| GP | General Practitioner |
| HESA | Higher Education Statistical Authority |
| HMRC | Her Majesty's Revenue and Customs |
| ICSO | Irish Central Statistical Office |
| IFD | Information Directorate |
| IHS | Integrated Household Survey |
| IPS | International Passenger Survey |
| LADUA | Local Authority District / Unitary Authority |
| LEA | Local Education Authority |
| LFS | Labour Force Survey |
| MYE | Mid Year Estimates |
| NHS | National Health Service |
| NINo | National Insurance Number |
| NIRS | National Insurance Recording System |
| NMD | New Migrant Databank |
| NMG | New Migrant Geography |
| NMGi | New Migrant Geography – immigration |
| NMGo | New Migrant Geography – emigration |
| NPD | National Pupil Dataset |
| NSOS | National Statistics Omnibus Survey |
| ONS | Office for National Statistics |
| PBS | Points Based System |
| PRDS | Patient Registration Data System |
| QNHS | Quarterly National Household Survey |
| SNPP | Sub-National Population Projections |
| TIM | Total International Migration |
| UK | United Kingdom |
| WRS | Workers Registration Scheme |

House of Lords Select Committee on Economic Affairs (2008b) The Economic Impact of Immigration Volume II: Evidence. 1st Report of Session 2007–08. HL Paper 82-II. http://www.publications.parliament.uk/pa/ld200708/ldselect/ldeconaf/82/82ii.pdf

Nowok, B., Kupiszewska, D. and Poulain, M. (2006) Statistics on international migration flows. Chapter 8, pp.203-231 in Poulain, M., Perrin, N. and Singleton, A. (eds.) THESIM: Towards Harmonised European Statistics on International Migration. Presses Universitaires de Louvain, Louvain.

Lanzieri, G. (2007) Long-term population projections at regional level. *Statistics in Focus*. 28/2007. Eurostat http://epp.eurostat.ec.europa.eu/cache/ITY_OFFPUB/KS-SF-07-028/EN/KS-SF-07-028-EN.PDF

NSCD (2006). *Estimating international migration for population estimates - an information paper*. National Statistics Centre for Demography. http://www.gro-scotland.gov.uk/files/uk-international-migration-paper.pdf.

ONS (2007a) Improved methods for estimating international migration – geographical distribution of estimates of in-migration. Titchfield: Office for National Statistics. http://www.statistics.gov.uk/downloads/theme_population/Geog_distn_in-migs.pdf

ONS (2007b) Improved methods for estimating international migration – geographical distribution of estimates of out-migration. Titchfield: Office for National Statistics. http://www.statistics.gov.uk/downloads/theme_population/Geog_distn_out-migs.pdf

ONS (2007c) Update on the development of the Integrated Household Survey. Titchfield: Office for National Statistics. http://www.ccsr.ac.uk/esds/events/2007-03-29/ihs/slides/bennett.ppt

ONS (2007d) A review of the potential use of administrative sources in the estimation of Population Statistics. Titchfield: Office for National Statistics. http://www.lga.gov.uk/lga/aio/1098388

ONS (2008a) *National Population Projections: 2006-based.* Series PP No 26. Basingstoke: Palgrave-Macmillan. http://www.statistics.gov.uk/downloads/theme_population/pp2no26.pdf

ONS (2008b) Updated Short-Term Migration Estimates, mid-2004 and mid-2005. Titchfield: Office for National Statistics. http://www.statistics.gov.uk/about/data/methodology/specific/population/future/imps/updates/downloads/STM_Update.pdf

ONS (2008c) *International Migration. Series MN No 33, 2006 Data.* Titchfield: Office for National Statistics. http://www.statistics.gov.uk/downloads/theme_population/MN33.pdf

ONS (2008d) Helping to Shape Tomorrow. The 2011 Census of Population and Housing in *England and Wales.* CM7513. London: Cabinet Office. http://www.ons.gov.uk/census/2011-census/2011-census-project/legislation/index.html

ONS (2008e) Population estimates for UK, England and Wales, Scotland and Northern Ireland - current datasets. 2006 Mid-year estimates. http://www.statistics.gov.uk/statbase/Product.asp?vlnk=15106

ONS (2008f) Improved Methods for Estimating International Migration: The Use of Calibration in Estimating International In-migration to UK Countries and the Regions of England.
18

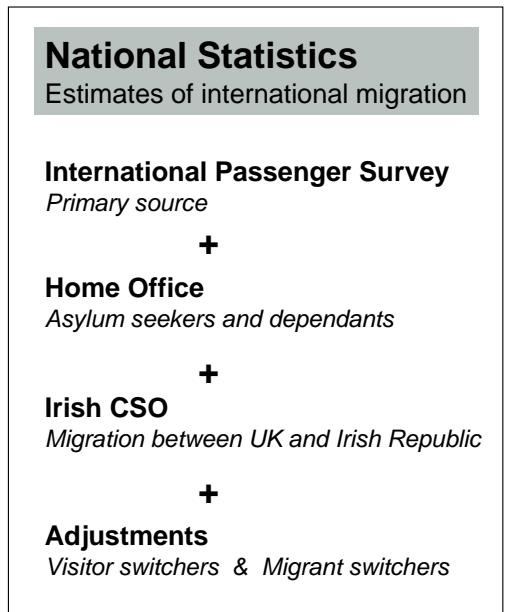

Source: ONS (2008c)

*Figure 1: The components that make up Total International Migration (TIM)*

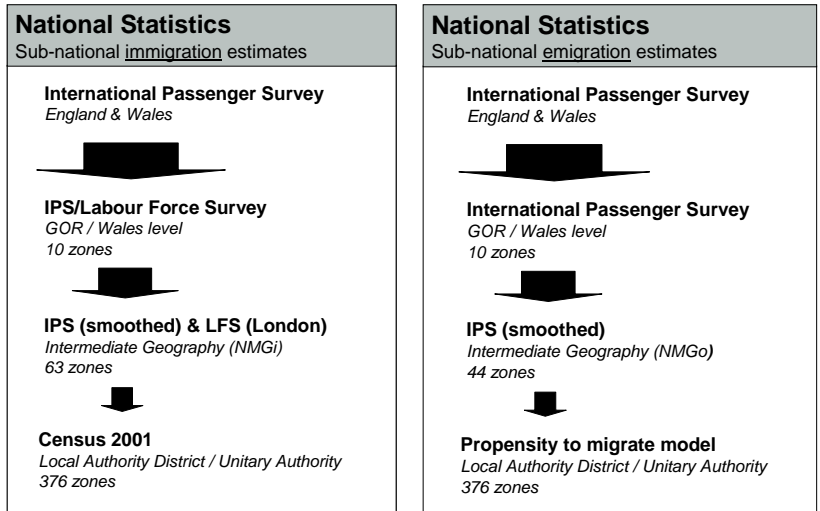

Source (ONS, 2008c)

*Figure 2: Total International Migration, sub-national estimation: methods employed*

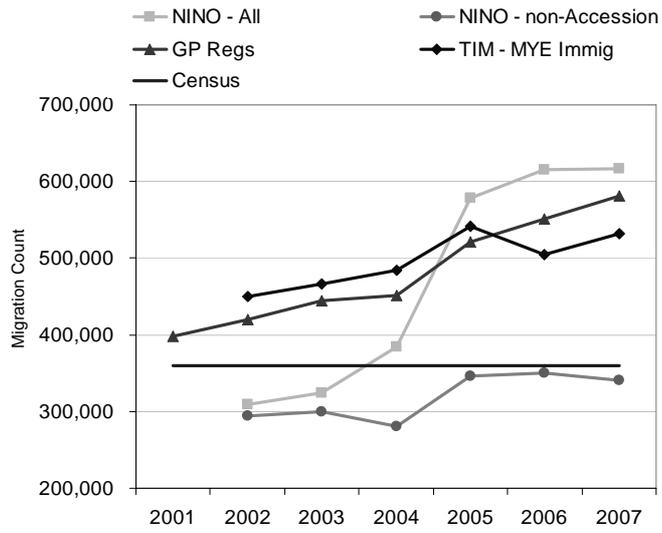

*Figure 3: Immigration trends, England, 2001-2007*

All data are Crown copyright. Sources: 100% data extract from the National Insurance Recording System (NIRS) with 2007/08 registrations recorded as 2007 in the charts: Components of mid-year population estimates (ONS, 2008e); GP registration statistics for calendar years provided by ONS

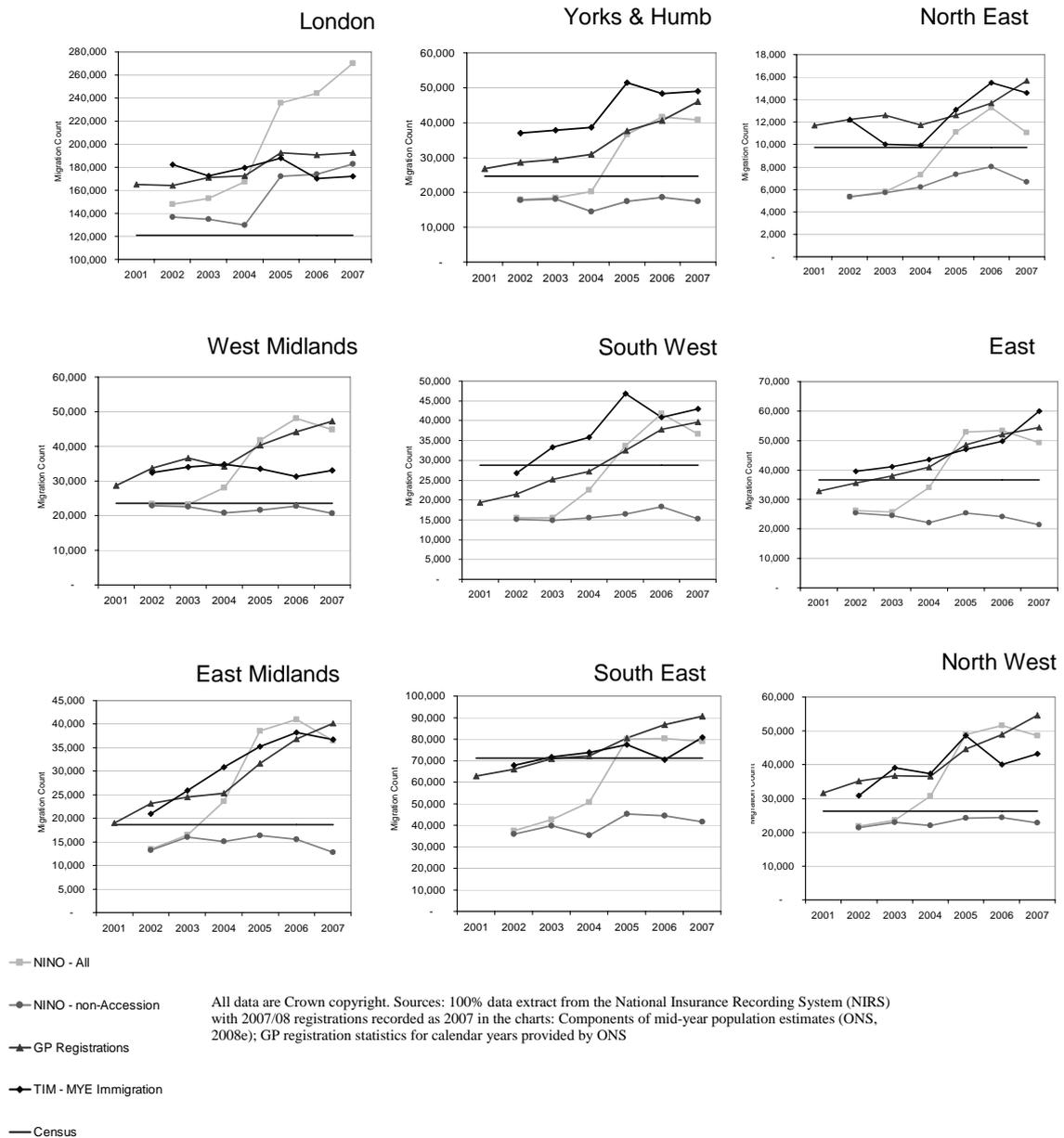

*Figure 4: Immigration trends, English Regions, 2001-2007*

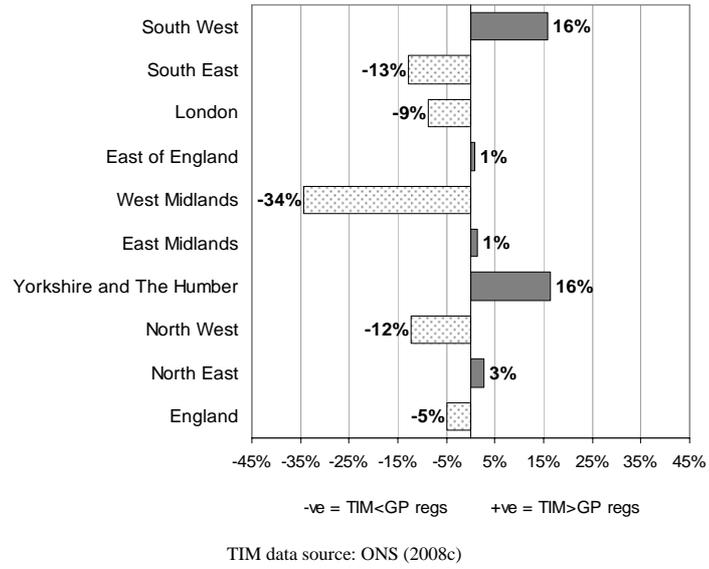

*Figure 5: Immigration totals, TIM versus GP registrations, 2005-2007*

|  | Data Sources | | |
|---|---|---|---|
|  | National | Regional distribution | Intermediate geography and local authority area distribution |
| Model A | TIM | GP Regs | GP Regs |
| Model B | TIM | GP Regs / NINo / HESA | GP Regs |

*Figure 6: Alternative models for estimating immigration*

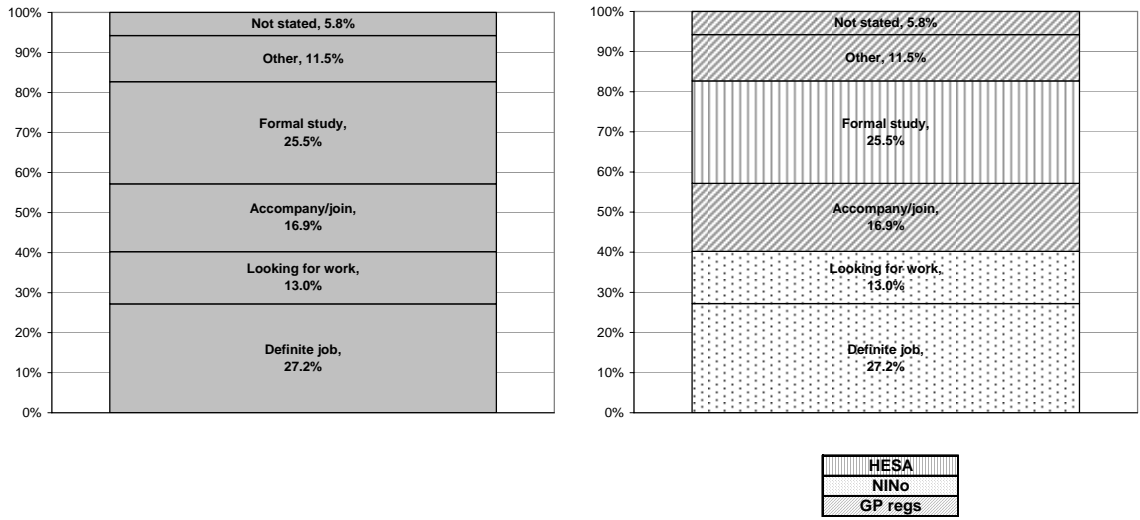

(a) TIM Reason profile 2004/06      (b) Datasets used to estimate immigration at regional level (GORs) in Model B

Source: ONS (2008c)

*Figure 7: Total International Migration, sub-national estimation*

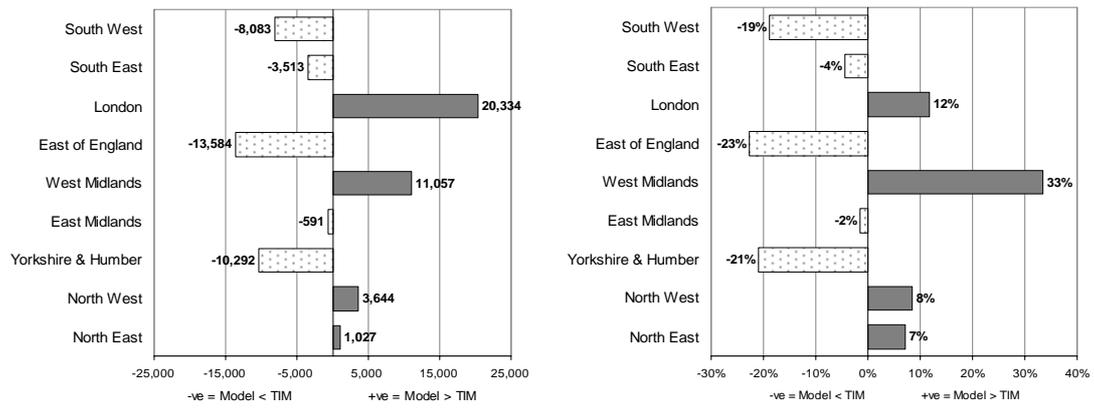

*Figure 8: Model B estimates minus ONS GOR allocation of TIM estimates: count and percentage differences, 2006*

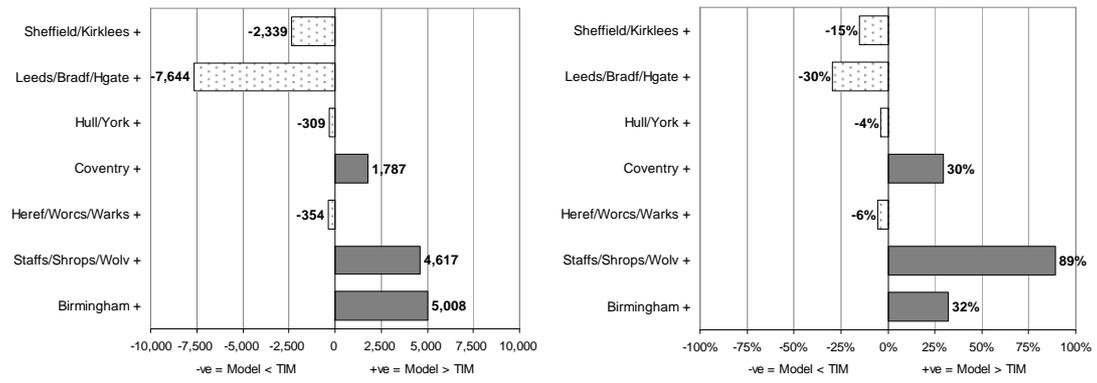

*Figure 9: Model B estimates minus ONS NMGi allocation of TIM estimates, Yorkshire and the Humber and the West Midlands, count and percentage differences, 2006*

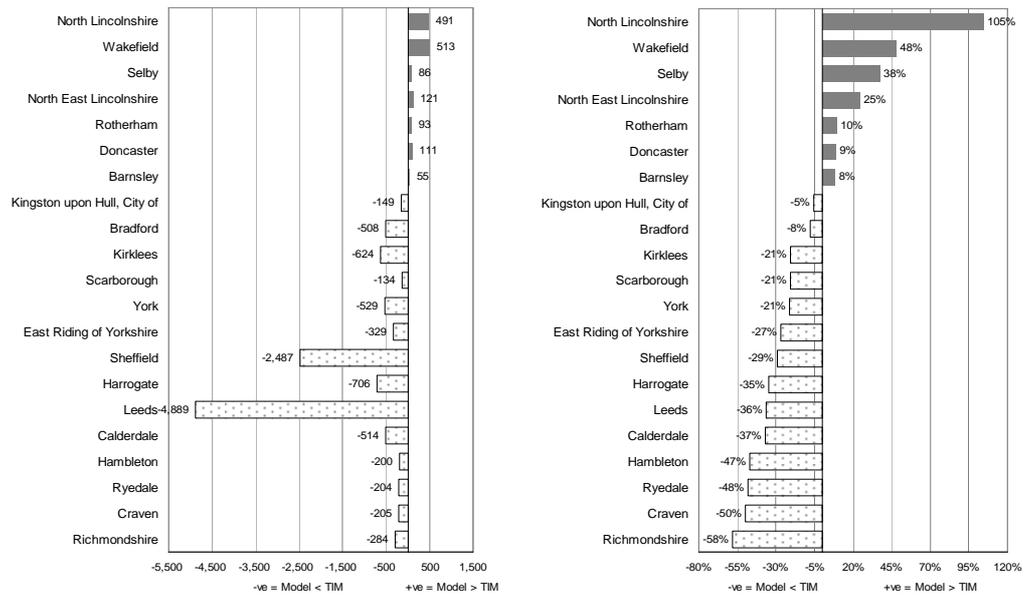

*Figure 10: Model B estimates minus ONS Local Authority allocation, Yorkshire and the Humber, count and percentage differences, 2006*

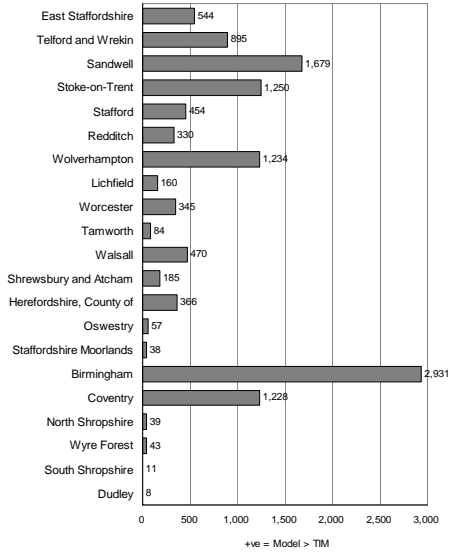
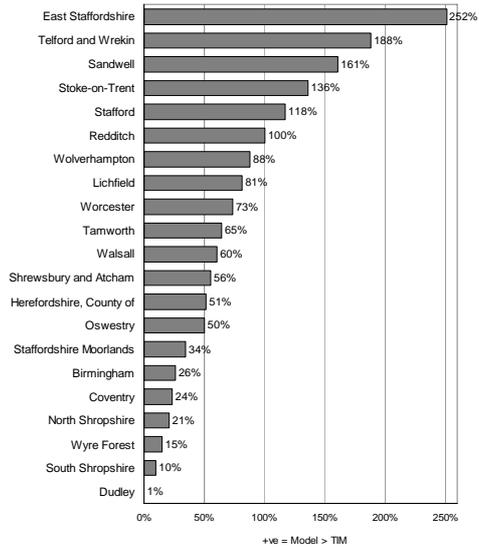
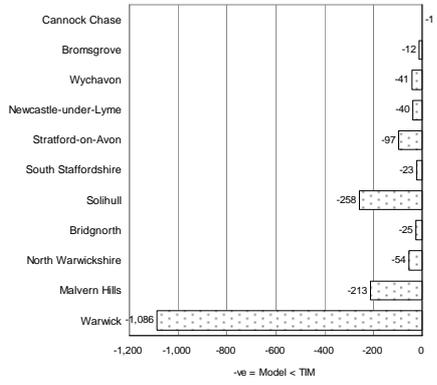
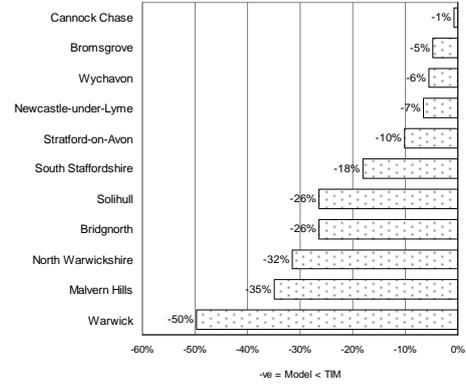

*Figure 11: Model B estimates minus ONS Local Authority allocation, West Midlands, count and percentage differences, 2006*

| Rank | Local authority | Model B > TIM estimate | | Rank | Local authority | Model B < TIM estimate | |
|---|---|---|---|---|---|---|---|
| | | Change | % | | | Change | % |
| 1 | Newham | 3,142 | 37% | 335 | South Northamptonshire | -932 | -75% |
| 2 | Birmingham | 2,931 | 26% | 336 | Southampton | -989 | -18% |
| 3 | Wandsworth | 2,810 | 39% | 337 | Southend-on-Sea | -1,011 | -48% |
| 4 | Lambeth | 2,767 | 45% | 338 | Charnwood | -1,047 | -40% |
| 5 | Hounslow | 2,556 | 44% | 339 | Portsmouth | -1,075 | -34% |
| 6 | Greenwich | 2,453 | 72% | 340 | Warwick | -1,086 | -50% |
| 7 | Ealing | 2,452 | 27% | 341 | Tower Hamlets | -1,308 | -17% |
| 8 | Waltham Forest | 2,256 | 54% | 342 | Plymouth | -1,335 | -42% |
| 9 | Barnet | 1,968 | 30% | 343 | Wokingham | -1,348 | -52% |
| 10 | Peterborough | 1,896 | 125% | 344 | Norwich | -1,378 | -37% |
| 11 | Lewisham | 1,858 | 44% | 345 | Bath and North East Somerset | -1,434 | -44% |
| 12 | Croydon | 1,841 | 42% | 346 | Forest Heath | -1,694 | -74% |
| 13 | Sandwell | 1,679 | 161% | 347 | Bristol, City of | -2,099 | -25% |
| 14 | Barking and Dagenham | 1,613 | 103% | 348 | Colchester | -2,318 | -53% |
| 15 | Hillingdon | 1,476 | 45% | 349 | Cambridge | -2,389 | -34% |
| 16 | Haringey | 1,464 | 25% | 350 | Sheffield | -2,487 | -29% |
| 17 | Bolton | 1,431 | 193% | 351 | Camden | -3,852 | -35% |
| 18 | Slough | 1,418 | 95% | 352 | Leeds | -4,889 | -36% |
| 19 | Stoke-on-Trent | 1,250 | 136% | 353 | Kensington and Chelsea | -5,128 | -55% |
| 20 | Wolverhampton | 1,234 | 88% | 354 | Westminster | -5,363 | -42% |

*Figure 12: Largest differences between ONS TIM estimates and Model B estimates, 2006*

*Table 1: Sources of data on UK international migration*

| Source | | | Coverage | | | Migrant Definition | | | | |
|---|---|---|---|---|---|---|---|---|---|---|
| Type and name | Respon-sibility | Reporting period | National | GOR | LADUA | Stocks | Flows | Migrant Population | Immigr-ation | Emigr-ation |
| **Censuses** | | | | | | | | | | |
| UK Census | ONS | Decennial | ✓ | ✓ | ✓ | ✓ | ✓ | All | ✓ | |
| Pupil Census | DCSF | Biannual | ✓ | ✓ | ✓ | ✓ | | Children joining State Schools | ✓ | |
| **Surveys** | | | | | | | | | | |
| International Passenger Survey (IPS) | ONS | Continuous | ✓ | ✓ | | | ✓ | All migrants | ✓ | ✓ |
| Labour Force Survey | ONS | Quarterly | ✓ | ✓ | | ✓ | ✓ | All (but with survey exclusions) | ✓ | |
| **Administrative datasets** | | | | | | | | | | |
| Immigration Control | HO | Continuous | ✓ | | | | ✓ | Non-EEA migrants | ✓ | |
| Work Permits | HO | Continuous | ✓ | ✓ | ✓ | | ✓ | Non-EEA migrants | ✓ | |
| National Insurance Number (NINo) Applications | DWP | Continuous | ✓ | ✓ | ✓ | | ✓ | Migrant workers | ✓ | |
| Worker Registration Scheme (WRS) | HO | Continuous | ✓ | ✓ | ✓ | | ✓ | A8 workers excluding self-employed | ✓ | |
| GP Registrations | NHS | Continuous | ✓ | ✓ | ✓ | | ✓ | All | ✓ | |
| Higher Education Statistics | HESA | Continuous | ✓ | ✓ | ✓ | ✓ | ✓ | Students | ✓ | |

Notes: See Glossary for the meaning of the acronyms

*Table 2: Immigration rates, TIM versus GP registrations, 2006*

| | Population mid-year 2006 | TIM Rate | GP Rate | Ratio TIM:GP |
|---|---|---|---|---|
| | 000s | migration per 1000 population | | % |
| East | 5,607 | 10.7 | 9.7 | 110% |
| South West | 5,124 | 8.4 | 7.7 | 108% |
| Yorkshire & Humber | 5,142 | 9.5 | 9.0 | 106% |
| North East | 2,556 | 5.7 | 6.1 | 93% |
| **ENGLAND** | **50,763** | **10.5** | **11.5** | **92%** |
| East Midlands | 4,364 | 8.4 | 9.2 | 91% |
| South East | 8,238 | 9.8 | 11.0 | 89% |
| London | 7,512 | 22.9 | 25.7 | 89% |
| North West | 6,853 | 6.3 | 8.0 | 79% |
| West Midlands | 5,367 | 6.2 | 8.8 | 70% |

Data are Crown copyright. Sources: 2006 Mid-year estimates (ONS, 2008e); TIM data (ONS, 2008c) GP registration statistics provided by ONS. Rate = ratio of immigrants to receiving region population.